\documentclass[11pt,prd,aps,epsfig,floats,preprint,eqsecnum,nofootinbib]{revtex4}
\pdfoutput=1
\usepackage{color}
\usepackage[squaren]{SIunits}
\usepackage{amsmath}
\usepackage{amsfonts}
\usepackage{slashed}
\usepackage{graphicx}
\usepackage{subfigure}
\usepackage{rotating}

\usepackage{graphicx,amsmath,amssymb,epsfig,verbatim,mathrsfs,array,layout,textcomp,latexsym}
\usepackage{multirow}

\allowdisplaybreaks[1]

\renewcommand{\(}{\left(}
\renewcommand{\)}{\right)}
\renewcommand{\{}{\left\lbrace}
\renewcommand{\}}{\right\rbrace}
\renewcommand{\[}{\left\lbrack}
\renewcommand{\]}{\right\rbrack}
\renewcommand{\Re}[1]{\mathrm{Re}\!\{#1\}}
\renewcommand{\Im}[1]{\mathrm{Im}\!\{#1\}}
\newcommand{\dd}[1][{}]{\mathrm{d}^{#1}\!\!\;}

\newcommand{\ie}{i.e.\,}
\newcommand{\cf}{cf.\,}
\newcommand{\refeq}[1]{Eq.~(\ref{eq:#1})}
\newcommand{\refeqs}[2]{Eqs.~(\ref{eq:#1})-(\ref{eq:#2})}
\newcommand{\reffig}[1]{Fig.~\ref{fig:#1}}
\newcommand{\refsec}[1]{Section \ref{sec:#1}}
\newcommand{\reftab}[1]{Table~\ref{tab:#1}}

\newcommand{\gfermi}{G_\mathrm{F}}
\newcommand{\GeV}{\,\mathrm{GeV}}

\newcommand{\wilson}[2][{}]{\mathcal{C}_{#2}^{\mathrm{#1}}}

\begin{document}

\setlength{\parindent}{0pt}

\vspace{-0.2cm}
\hbox{DO-TH 10/16}

\title{More Benefits of Semileptonic Rare $B$ Decays at Low Recoil:\\ CP Violation}

\author{Christoph Bobeth} \affiliation{
Institute for Advanced Study \& Excellence Cluster Universe, Technische Universit\"at M\"unchen,
   D-85748 Garching, Germany}
\author{Gudrun Hiller} \author{Danny van Dyk} \affiliation{Institut f\"ur
  Physik, Technische Universit\"at Dortmund, D-44221 Dortmund, Germany}

\begin{abstract} 
  We present a systematic analysis of the angular distribution of
  $\bar{B}\to\bar{K}^* (\to \bar{K}\pi) l^+l^-$ decays with $l=e,\mu$ in the low
  recoil region (\ie at high dilepton invariant masses of the order of the mass of the $b$-quark) to 
  account model-independently  for CP violation beyond the Standard Model, working to
  next-to-leading order QCD.
 {}From the employed heavy quark effective theory framework we identify
  the key CP observables with reduced hadronic uncertainties. 
  Since some of the CP asymmetries
 are CP-odd they  can be measured without $B$-flavour tagging.
 This is particularly beneficial for 
  $\bar{B}_s, B_s \to \phi (\to K^+ K^-) l^+l^-$ decays, which are not self-tagging, and we work
  out the corresponding time-integrated CP asymmetries.
Presently available experimental constraints allow  the proposed CP asymmetries to be sizeable, up to values of the order  $\sim 0.2$,  while the corresponding Standard Model values receive a strong 
parametric suppression at the level of  ${\cal{O}}(10^{-4})$. Furthermore, we work out the allowed ranges of  the short-distance (Wilson) coefficients $\wilson{9,10}$ in the presence of  CP violation beyond the Standard Model but no further Dirac structures.
We find the  $\bar{B}_s \to \mu^+\mu^-$ branching ratio to be below $9 \times 10^{-9}$ 
(at  95\% CL).
 Possibilities to check the performance of the theoretical low recoil framework
are pointed out.
    \end{abstract}

\maketitle

%
%
\section{Introduction}

The exclusive rare flavour changing neutral current (FCNC) decay
$\bar{B}\to\bar{K}^* (\to \bar{K}\pi) l^+l^-$ with $l = e, \mu$ has high
sensitivity to physics
beyond the Standard Model (BSM) due to the large number of complementary
measurements possible from the full angular distribution  \cite{Kruger:1999xa}.
Many works have focussed 
on the region of low dilepton invariant mass squared, $q^2$, typically taken within the range 
$1$--$6\GeV^2$. The latter is accessible to QCD factorisation
\cite{Beneke:2001at, Beneke:2004dp}, which has enabled systematic studies of CP-averaged observables as well as
CP-asymmetries \cite{Bobeth:2008ij, Egede:2008uy, Altmannshofer:2008dz,
  Bobeth:2009ku, Egede:2010zc}.
Intermediate values of $q^2$ fall into the
narrow-resonance region dominated by the pronounced $c\bar{c}$-resonance background 
from the decays $\bar{B} \to \bar{K}^* \{J/\psi, \psi'\} \to \bar{K}^* l^+l^-$, recently  studied 
in \cite{KMPW} including  also low $q^2$ tails.
At larger dilepton masses,
at about $q^2 \gtrsim 14 $~GeV${}^2$, follows the broad resonance region. The
latter is characterised by the low recoil of the hadronic system.  Here, the large values of $q^2
\sim m_b^2$, where $m_b$ denotes the mass of the $b$-quark, allow to perform an operator product expansion (OPE) \cite{Grinstein:2004vb,Beylich:2011aq} which, when combined 
with heavy quark effective theory (HQET) and the corresponding heavy quark form factor relations
\cite{Grinstein:2002cz}, leads to powerful predictions, see \cite{Grinstein:2004vb} and 
Hurth and Wyler in \cite{Hewett:2004tv}.

In fact, it has been shown recently that the heavy quark framework 
applied to the low recoil region results in a very simple amplitude structure of the decays
$\bar{B}\to\bar{K}^* (\to \bar{K}\pi) l^+l^-$   \cite{Bobeth:2010wg}.
 Specifically, in the heavy quark limit, all three participating transversity
amplitudes obey
\begin{align}
  \label{eq:Aloreco}
  A_i^{L,R} & \propto C^{L,R} \times f_i\,, & 
  i & = \perp,\parallel,0 \, ,
\end{align}
hence factorise into universal short-distance coefficients $C^{L,R}$ and 
form factor coefficients $f_i$.  
This feature can be greatly exploited  to enhance the BSM sensitivity, to test form factor predictions against data and to check the goodness of the OPE framework. More explicit,
the angular distribution of $\bar{B}\to\bar{K}^* (\to \bar{K}\pi) l^+l^-$ decays 
allows for observables with the following salient properties, see \cite{Bobeth:2010wg}
for details:

\bigskip

$i)$ The observable $H_T^{(1)} = 1$ does $not$ depend on
short-distance coefficients $nor$ on form factors. 

$ii)$ The observables 
$H_T^{(2,3)}$ depend on the short-distance coefficients only, and obey $H_T^{(2)}=H_T^{(3)}$.

$iii)$ Several observables can be formed
 which depend on the form factors only. 

$iv)$ The 
angular observables $J_{7,8,9}$,  which are odd under naive time-reversal, vanish.

\bigskip 

Beyond zeroth order in $1/m_b$, the influence of the power corrections 
is weak because the $\Lambda_{\rm QCD}/m_b$ corrections are parametrically suppressed:
The ones to the form factor relations 
enter with a suppression by small ratios of Wilson coefficients 
and the ones from subleading operators in the OPE arise at ${\cal O}(\alpha_s)$ only. Moreover,
the relevant hadronic matrix elements from both sources are not independent \cite{Grinstein:2004vb}. 
While the latter matrix elements 
are currently not known from first principles for $B \to K^*$, model
estimates suggest that they are at least not enhanced beyond the naive expectations~\cite{Grinstein:2002cz}.

In this paper we extend previous works \cite{Bobeth:2010wg} on the low recoil region by allowing for BSM CP violation.
We work to next-to-leading order (NLO) in QCD and to lowest order (LO) in 
$1/m_b$. The ${\cal{O}}(\Lambda_{\rm QCD}/m_b)$ corrections  are taken into account in the
estimation of  the uncertainties. 
Further higher order corrections, including charm loops  with gluons are power-suppressed
at low recoil 
\cite{Grinstein:2004vb,Beylich:2011aq} and not considered given the targeted precision.
The consistency between the outcome of an analysis excluding and using only the high-$q^2$ region data \cite{Bobeth:2010wg} supports the employed OPE framework.

We propose and study CP observables with only subleading form factor uncertainties 
in Sections \ref{sec:cpv-lorecoil} and \ref{sec:cp-asym-lorecoil}. In Section
\ref{sec:untagged-cp-asym} we calculate mixing-induced
time-integrated CP asymmetries relevant for  the decays
$B_s, \bar{B}_s\to\phi(\to K^+K^-)\,l^+l^-$.
In Section \ref{sec:aCP:rel:obs} we give the relations between the CP observables and the angular distributions
in $\bar{B}\to \bar{K}^*l^+l^-$ or likewise $\bar{B}_s\to\phi\,l^+l^-$ decays. We work out the
constraints on the complex-valued short-distance coefficients  in Section \ref{sec:constraints} and summarize
in Section \ref{sec:conclusion}. In an appendix we present the method used to
estimate the uncertainties from the  ${\cal{O}}(\Lambda_{\rm QCD}/m_b)$ corrections.

%
%
\section{Low Recoil CP Asymmetries}

Our aim is to extend our previous study of $\Delta B=1$ radiative and
semileptonic decays \cite{Bobeth:2010wg} in the presence of CP violation. 
We use a model-independent framework with an effective Hamiltonian
\begin{align}
  \nonumber
  {\cal{H}}_{\rm{eff}} & = 
  - \frac{\gfermi}{\sqrt{2}} \frac{e}{4 \pi^2} V_{tb}^{} V^*_{ts} 
  \Big( e\, {\cal C}_9 \[\bar{s} \gamma_\mu P_{L} b\]\!\[\bar{l} \gamma^\mu l\] 
      + e\, {\cal C}_{10} \[\bar{s} \gamma_\mu P_{L} b\]\!\[\bar{l} \gamma^\mu \gamma_5 l\] 
\\
  \label{eq:Heff} 
   & \hspace{3.5cm} + {\cal C}_7 m_b \[\bar{s} \sigma^{\mu\nu} P_{R} b\]\!F_{\mu\nu} \Big) 
     + {\rm h.c.} + \ldots \,,
\end{align}
where the ellipses denote contributions which we assume to be SM-like because
they are either subdominant in the radiative and semileptonic $b \to s$
decay amplitudes or  induced in the SM at tree level. With CP violation
beyond the SM, the Wilson coefficients ${\cal{C}}_7$, ${\cal{C}}_9$ and
${\cal{C}}_{10}$ are complex-valued. All other Wilson coefficients are assumed
to be SM-like, and are real-valued after factoring out the Cabibbo-Kobayashi-Maskawa
(CKM) factors $V_{ik}$, similar to \refeq{Heff}.  For details and definitions we refer to
\cite{Bobeth:2010wg}, which we follow closely. In particular, we take all numerical input as in
\cite{Bobeth:2010wg} except for the CKM one, which we calculate 
from the Wolfenstein parameters $A=0.812^{+0.013}_{-0.027}$,
$\lambda=0.22543\pm0.00077$, $\bar \rho=0.144\pm{0.025}$ and
$\bar \eta=0.342^{+0.016}_{-0.015}$ \cite{Charles:2004jd}.
In the following we understand all Wilson coefficients to be evaluated at the
scale $\mu_b \approx m_b$. The SM values of the most important ones are
approximately, to next-to-next-to-leading logarithmic (NNLL) order,
\begin{align} \label{eq:SMvalues}
  \mathcal{C}_7^{\rm SM} & = -0.3 , &
  \mathcal{C}_9^{\rm SM} & = 4.2 , &
  \mathcal{C}_{10}^{\rm SM} & = -4.2\,.
\end{align}

%
\subsection{Decay amplitudes with CP violation at low recoil}
\label{sec:cpv-lorecoil}

In a previous work \cite{Bobeth:2010wg} we identified
\begin{align} 
  \label{eq:rho1:def}
  \rho_1 & = \frac{1}{2} \left(|C^R|^2 + |C^L|^2\right) =
  \left|\wilson[eff]{9} + \kappa \frac{2\hat{m}_b}{\hat{s}}\wilson[eff]{7}\right|^2 
  + \left|\wilson{10}\right|^2 , 
\\
  \rho_2 & = \frac{1}{4} \left(|C^R|^2 - |C^L|^2\right) = 
  \Re{\(\wilson[eff]{9} 
  + \kappa \frac{2\hat{m}_b}{\hat{s}} \wilson[eff]{7}\) \wilson[*]{10}} ,
\end{align}
as the only independent short-distance factors which enter the observables of 
the angular distribution of $\bar B \to \bar K^* (\to \bar{K} \pi) l^+ l^-$
decays at low hadronic recoil. Here, $\hat s =q^2/m_B^2$ and $\hat m_b=m_b/m_B$,
where $m_B$ denotes the mass of the $B$ meson.
The factor $\kappa$ accounts for the relation between the dipole ($T_{1,2,3}$) and
(axial-) vector ($V,A_{1,2}$) form factors that can be calculated systematically
\cite{Grinstein:2004vb,Grinstein:2002cz}.  At lowest order in $1/m_b$ and 
including ${\cal{O}}(\alpha_s)$ corrections, it reads
\begin{align}
  \kappa & =
  1 - 2\, \frac{\alpha_s}{3\pi} \ln \left(\frac{ \mu}{m_b}\right) .
\end{align}
The effective coefficients are written as
\begin{align}
  \label{eq:c7effGP}
  \wilson[eff]{7} & = 
  \wilson{7} - 
  \frac{1}{3} \[ \wilson{3} + \frac{4}{3}\,\wilson{4} + 20\,\wilson{5} 
    + \frac{80}{3}\wilson{6} \]
    + \frac{\alpha_s}{4 \pi} \[ \left(\wilson{1} - 6\,\wilson{2}\right) A(q^2) 
    - \wilson{8} F_8^{(7)}(q^2)\] ,
\end{align}
and
\begin{align}  
  \label{eq:c9effGP}
  \wilson[eff]{9} & = 
    \wilson{9} + 
    h(0, q^2) \[ \frac{4}{3}\, \wilson{1} + \wilson{2} + \frac{11}{2}\, \wilson{3}
    - \frac{2}{3}\, \wilson{4} + 52\, \wilson{5} - \frac{32}{3}\, \wilson{6}\] 
\\
  & - \frac{1}{2}\, h(m_b, q^2) \[ 7\, \wilson{3} + \frac{4}{3}\, \wilson{4} + 76\, \wilson{5}
    + \frac{64}{3}\, \wilson{6} \]
    + \frac{4}{3} \[ \wilson{3} + \frac{16}{3}\, \wilson{5} + \frac{16}{9}\, \wilson{6} \]
\nonumber\\[0.2cm]
  & + \frac{\alpha_s}{4 \pi} \[ \wilson{1} \left(B(q^2) + 4\, C(q^2)\right) 
    - 3\, \wilson{2}\left(2\, B(q^2) - C(q^2)\right) - \wilson{8}F_8^{(9)}(q^2) \]  
\nonumber\\[0.2cm] 
  & + 8\, {\frac{m_c^2}{q^2}  \[ \left( \frac{4}{9}\,\wilson{1} 
    + \frac{1}{3}\,\wilson{2} \right)(1+\hat \lambda_u) + 2\,\wilson{3} + 20\,\wilson{5} \] },  
\nonumber 
\end{align}
where we extended previous works \cite{Grinstein:2004vb, Bobeth:2010wg} by including the doubly
Cabibbo-suppressed contribution proportional to $\hat\lambda_u =
V_{ub}^{}V_{us}^*/(V_{tb}^{}V_{ts}^*)$.  The latter is responsible for CP
violation in the SM and appears only in the coefficient $ \wilson[eff]{9}$ with
$m_c^2/q^2$ suppression. We refer to \cite{Grinstein:2004vb, Bobeth:2010wg} for
more details concerning the (real-valued) Wilson coefficients $\wilson{i\leq 6}$ as
well as the LO and NLO QCD corrections encoded in the functions $h(m_i, q^2)$ and $A,B,C$, $F_8^{(7,9)}$, respectively.

In the presence of CP violation, there are  four independent short-distance factors
\begin{align}
  \label{eq:rho:bar}
  \rho_1, \rho_2 ~~~\mbox{and}~~~ \bar \rho_1,\bar \rho_2 \, ,
\end{align}
where the barred factors are obtained from the unbarred ones by complex conjugation of  the Wilson coefficients ${\cal{C}}_i$ and the CKM factor
$\hat\lambda_u$. The building blocks describing CP violation are hence
\begin{align} 
  \Delta \rho_i & = \rho_i-\bar \rho_i ~~~\mbox{with}~~~ i = 1, 2 \,.
\end{align}
They can be written as
\begin{align} 
  \label{eq:delro1}
  \Delta\rho_1 & = 
  4 \, {\rm Im}\,Y \cdot \Im{ \wilson{9} + \kappa \frac{2 \hat m_b}{\hat s}
  \wilson{7} +  Y_9^{(u)} \hat{\lambda}_u } \, ,
\\
  \label{eq:delro2}
  \Delta\rho_2 & = 
  2 \, {\rm Im}\,Y \cdot { \rm Im} \, \wilson{10} \, , 
\end{align}
using
\begin{align}
  Y & = Y_9 + \kappa \frac{2 \hat m_b}{\hat s} Y_7
\end{align}
and the decomposition of Eqs.~(\ref{eq:c7effGP}) and (\ref{eq:c9effGP}) into
\begin{align} \label{eq:decomp}
  \wilson[eff]{7} & = \wilson{7} + Y_7\, , &
  \wilson[eff]{9} & = \wilson{9} + Y_9 + \hat\lambda_u Y_9^{(u)} \, .
\end{align}
Note that $Y_9^{(u)}$ is real-valued, and we suppress the $q^2$-dependence in
the effective coefficients and the $Y_i$ throughout this work. 
It follows from \refeqs{delro1}{delro2} 
that $\Delta \rho_1$ probes
the weak phases of $\wilson{7}$ and $\wilson{9}$, whereas $\Delta \rho_2$ probes
the weak phase of $\wilson{10}$. The imaginary parts of the $Y_i$  give rise to the strong phases and hence drive the magnitude of CP violation.

\begin{figure}
\begin{center}
  \includegraphics[height=.25\textheight]{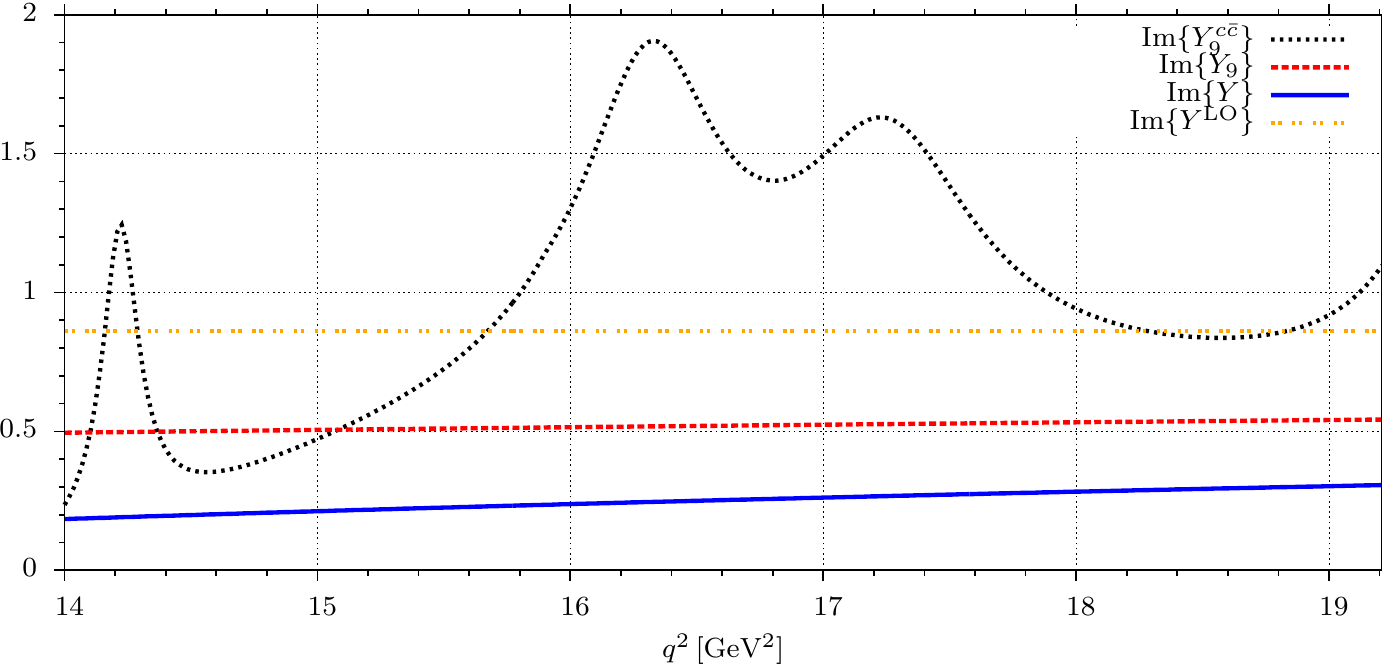}
\end{center}
\caption{The imaginary parts of $Y$ (solid, blue) and $Y_9$ (dashed, red) in the OPE
\cite{Grinstein:2004vb}   including
NLO $\alpha_s$-corrections as functions of~$q^2$ in the low recoil region. The LO result, where ${\rm Im}\, Y = {\rm Im}\, Y_9$ is shown by the dashed-dotted (orange) curve.
  The $c\bar{c}$-resonance curve (dotted, black) shows the imaginary part of 
  $Y_9$ from $e^+ e^- \to hadrons$ data~\cite{Kruger:1996cv,Nakamura:2010zzi}. 
  \label{fig:strong-phase}}
\end{figure}

In   \reffig{strong-phase}  we show the imaginary
part of $Y_9$ (dashed curve) and $Y$ (solid curve) from the OPE \cite{Grinstein:2004vb} at NLO  QCD using Eqs.~(\ref{eq:c7effGP}) and (\ref{eq:c9effGP}).
The NLO QCD corrections to both ${\rm Im}\, Y_{7,9}$ are sizeable and lead to a
reduction of the strong phases compared to the LO value of ${\rm Im}\, Y$ {(dashed-dotted curve)}. 
Since ${\rm Im} \, Y_7$ vanishes at LO, the NLO corrections constitute the leading
contribution to this quantity. 

Also shown in \reffig{strong-phase} is the absorptive part ${\rm Im} \, Y^{c \bar c}$ (dotted curve) obtained from a phenomenological fit to
$e^+ e^- \to hadrons$ data assuming factorization \cite{Kruger:1996cv}. The Breit-Wigner amplitude matches the charmonium peaks  of the branching ratios
$\mathcal{B}(\bar{B}\to\bar{K}^* c\bar{c})$ for $c \bar c =J/\Psi$
and $\Psi^\prime$. Note that  for NNLL
values of the Wilson coefficients $\wilson{1}$ and $\wilson{2}$ and with present day data
\cite{Nakamura:2010zzi} no fudge factor \cite{Ali:1999mm} is needed. 
We apply this ansatz for  the higher $c\bar{c}$-resonances as well, where presently
no $B$-decay data exist. The fit exhibits the local charm resonance structure
from $\bar B \to \bar K^* (c \bar c) \to \bar K^* l^+ l^-$ decays.  As can be seen in  \reffig{strong-phase}, the resonance contribution is of the same order as the OPE prediction at LO QCD
and indicates comparable results after integrating over a sufficiently large
region in the dilepton mass. However, we find a factor of $\sim\!3$ between both
approaches when using NLO QCD corrections and integrating over the low recoil region.

%
\subsection{The low recoil CP asymmetries}
\label{sec:cp-asym-lorecoil}

At low recoil and within our framework (LO in $1/m_b$, SM operator basis \refeq{Heff} ) all $\bar B \to
\bar K^* l^+ l^-$ observables are, as far as short-distance physics is
concerned, either proportional to $\rho_1$, $\rho_2/\rho_1$ or short-distance
insensitive \cite{Bobeth:2010wg}. Consequently, there are only two types of CP
asymmetries:
\begin{align} 
  \label{eq:acp1and2}
  a_{\rm CP}^{(1)} & \equiv 
    \frac{\rho_1 - \bar\rho_1}{\rho_1 + \bar\rho_1}\,, &
  a_{\rm CP}^{(2)} & \equiv  
    \frac{ \frac{\rho_2}{\rho_1} - \frac{\bar\rho_2}{\bar\rho_1}}
           {\frac{\rho_2}{\rho_1} + \frac{\bar\rho_2}{\bar\rho_1}}\,.
\end{align}
It is advantageous to define further
\begin{align}
  \label{eq:acp3}
  a_{\rm CP}^{(3)} & \equiv 
    2\, \frac{\rho_2 - \bar\rho_2}{\rho_1 + \bar\rho_1}\,,
\end{align}
which is not independent of $a_{\rm CP}^{(1,2)}$. Since $(\rho_1 +
\bar{\rho}_1)$ is positive definite, see \refeq{rho1:def}, it is in general
better suited for normalisation then the denominator of $a_{\rm CP}^{(2)}$, which
might cross zero and could make the  theoretical uncertainties blow up. We note
that $a_{\rm CP}^{(1)}$ equals the direct CP asymmetry in the rate, $A_{\rm
  CP} \equiv \frac{\Gamma-\bar\Gamma}{\Gamma+\bar\Gamma}$. Furthermore, $a_{\rm
  CP}^{(2)} $ equals  $A_{\rm FB}^{\rm CP} \equiv \frac{A_{\rm FB}+\bar A_{\rm
    FB}}{A_{\rm FB}-\bar A_{\rm FB}}$ \cite{Buchalla:2000sk}, the CP
asymmetry of the forward-backward asymmetry, whereas $a_{\rm CP}^{(3)}$ corresponds
to the low recoil transversity observables $H_T^{(2,3)}$
 introduced in Ref.~\cite{Bobeth:2010wg}. No further CP asymmetries
can be formed from the decays at low recoil beyond \refeqs{acp1and2}{acp3}
unless one considers neutral meson mixing, which we do in the next section. Note
that $a_{\rm CP}^{(1,2)}$ are related to observables which require $B$-flavour
tagging, whereas $a_{\rm CP}^{(3)}$  can be
extracted from untagged $B$ meson samples [\cf \refsec{aCP:rel:obs}\,].

Using $\Delta\rho_i \ll \rho_i, \bar{\rho}_i$ the expressions for the CP
asymmetries simplify to
\begin{align} 
  \label{eq:approx1}
  a_{\rm CP}^{(1)}
    & \approx \frac{\Delta\rho_1}{2\rho_1} \,, &
  a_{\rm CP}^{(2)}
    & \approx \frac{\Delta\rho_2}{2\rho_2} - a_{\rm CP}^{(1)} \,,&
  a_{\rm CP}^{(3)}
    & \approx \frac{\Delta\rho_2}{\rho_1} \,.
\end{align}

The SM values of the CP asymmetries are induced at the order $(m_c^2/m_b^2) \,
{\rm Im} \hat \lambda_u \sim 10^{-3}$ and are tiny.  Since the CP asymmetries at
low recoil are T-even only, a finite strong phase is needed for a finite CP
asymmetry. The strong phase is roughly given as $\arg(Y)$, yielding an additional
suppression by another order of magnitude. Therefore, at low recoil
\begin{align}
  \left|a_{\rm CP}^{(1,2,3)}\right|_{\rm SM}
    & \lesssim  10^{-4} \, .
\end{align}
Given the foreseen experimental precision, the
CP asymmetries in the SM are therefore completely negligible due to their strong
parametric suppression. Hence, any observed finite CP asymmetry is a signal of
physics beyond the SM.

Beyond the SM, the CP asymmetries at low recoil can be significantly enhanced.
To estimate the order of magnitude of $a_{\rm CP}^{(1,2,3)}$ we assume that
$|{\cal{C}}_{7,9,10}|$ are close to their respective SM values where
$|\wilson[SM]{7}| \ll |\wilson[SM]{9,10}|$.  Then, using \refeq{approx1} and
${\rm Im} \, Y$ as shown in \reffig{strong-phase}, we obtain, roughly,
\begin{align} 
  \label{eq:acp-np1}
  a_{\rm CP}^{(1)} & \simeq 
    \frac{2\, {\rm Im}  Y }{|\wilson{9}|^2 +|\wilson{10}|^2} 
    {\rm Im} \left(\wilson{9} + \kappa \frac{2 \hat m_b}{\hat s} \wilson{7} \right)
    \lesssim \mathcal{O}(0.1)\,,
\\[0.2cm]
   \label{eq:acp-np2}
  a_{\rm CP}^{(2)} & \simeq
    \frac{ {\rm Im} Y {\rm Im}\, \wilson{10}}{{\rm Re} \, \wilson{9} \wilson[*]{10}} 
    -a_{\rm CP}^{(1)} \lesssim \mathcal{O}(0.1)\,,
\\[0.2cm]
   \label{eq:acp-np3}
  a_{\rm CP}^{(3)} & \simeq 
    \frac{ {\rm Im} Y {\rm Im}\, \wilson{10}}{|\wilson{9}|^2 +|\wilson{10}|^2}
    \lesssim \mathcal{O}(0.1)\,,
\end{align}
in agreement with \cite{Kruger:2000zg,Buchalla:2000sk,Alok:2011gv}.
Note that for very small BSM values of ${\rm Re} \, \wilson{9} \wilson[*]{10}$ 
the values of $a_{\rm CP}^{(2)}$ become unconstrained.
Furthermore, for  ${\rm Im}\, \wilson{10} = 0$ the following relations hold
\begin{align}
  a_{\rm CP}^{(1)} & = -a_{\rm CP}^{(2)}\,, &
  a_{\rm CP}^{(3)} & = 0\,.
\end{align}

To investigate more quantitatively  the above CP asymmetries we define a BSM benchmark point
\begin{align}
  \wilson{7} & = -0.3\, i \,,& 
  \wilson{9} & = +4.2\, i \,,& 
  \wilson{10} = -4.2\, i\,,
  \label{eq:benchmark}
\end{align}
which passes all the current experimental constraints. In particular, both
interference terms ${\rm Re} \, \wilson{9} \wilson[*]{10}$  and ${\rm Re} \, \wilson{9} \wilson[*]{7}$
which are probed by $A_{\rm FB}(\bar B \to \{ X_s,\bar{K}^{*} \} l^+ l^-)$ and the 
$\bar B \to \{ X_s,\bar{K}^{(*)} \} l^+ l^-$ branching ratios, respectively, are SM-like.
Due to the 
maximal phases the benchmark values induce large BSM  CP violation. 
\begin{table}
\begin{tabular}{c||c|c|c}
    & LO & NLO & $\quad$ $c\bar{c}$-resonances $\quad$
\\
\hline \hline &&
\\[-0.5cm]
  $\quad \langle a_{\rm CP}^{(1)} \rangle \quad$ & 
  $\quad + 0.208\, {^{+0.036}_{-0.050}}\Big|_{\rm SD}\, {\pm 0.006}\Big|_{\rm
    SL} \quad$ & 
  $\quad + 0.074\, {^{+0.029}_{-0.037}}\Big|_{\rm SD}\, {\pm 0.011}\Big|_{\rm
    SL} \quad$ &
  $\quad + 0.17 \quad$
\\[0.3cm]
  $\langle a_{\rm CP}^{(2)} \rangle$ &
  $ + 0.071\, {^{+0.016}_{-0.015}}\Big|_{\rm SD}\, {\pm 0.009}\Big|_{\rm SL}$ & 
  $ + 0.020\, {^{+0.005}_{-0.008}}\Big|_{\rm SD}\, {\pm 0.006}\Big|_{\rm SL}$ &
  $ + 0.06$
\\[0.3cm]
  $\langle a_{\rm CP}^{(3)} \rangle$ &
  $ - 0.257\, {^{+0.056}_{-0.035}}\Big|_{\rm SD}\, {\pm 0.009}\Big|_{\rm SL}$ & 
  $ - 0.090\, {^{+0.044}_{-0.031}}\Big|_{\rm SD}\, {\pm 0.010}\Big|_{\rm SL}$ &
  $ - 0.20$
\end{tabular}
\caption{ \label{tab:aCP:bench}
  The $q^2$-integrated low recoil CP asymmetries 
  at LO and NLO QCD for the BSM benchmark point (\cf~\refeq{benchmark}) with the main uncertainties from the variation of the renormalisation scale 
 $\mu_b$ (SD) and subleading power corrections 
  (SL). Also given are the asymmetries using $c\bar{c}$-resonance data~\cite{Kruger:1996cv,Nakamura:2010zzi}.
}
\end{table}
The CP asymmetries evaluated at the benchmark point 
are given in \reftab{aCP:bench}. Throughout this work we use $\langle .. \rangle$ to denote the
integrated observables formed out of integrated angular coefficients following Ref.~\cite{Bobeth:2010wg}. For  the low recoil integration region we take $14 \GeV^2 < q^2 \leq 19.2 \GeV^2$.
We find that the main parametric
uncertainty in $\langle a_{\rm CP}^{(i)} \rangle$ stems from the variation of the renormalisation scale $\mu_b \in [2.1, 8.4]$~GeV (SD).  When varying the scale $\mu_b$, the model-independent BSM contributions to the Wilson coefficients \refeq{benchmark} are assumed to be
given at the reference scale $\mu = 4.2$~GeV and their renormalisation 
group evolution to $\mu_b$ is taken into account in leading logarithmic approximation. 
While an efficient cancellation of the $\mu_b$-dependence is at work in the sum
${\cal C}_i^{\rm eff}={\cal C}_i+Y_i$, the numerators of the CP asymmetries
depend on the product of ${\rm Im} \, {\cal C}_i$ and  ${\rm Im} Y$, and result in the large
reported uncertainties.
The subleading corrections
to the Isgur-Wise form factor relations and transversity amplitudes (SL)  constitute
another major source of uncertainty.  Its  estimate is explained in Appendix
\ref{app:sub-lead:crr}.

As can be seen from   \reftab{aCP:bench} the impact of the NLO corrections is sizable
on the CP asymmetries. The LO predictions are
about a factor 3 larger than the NLO ones due to large destructive NLO
contributions to Im$\,Y$. In fact, concerning Im$\,Y_7$ the NLO corrections
constitute the leading contribution which also implies a large scale uncertainty
at NLO, but the NLO corrections are sizeable in Im$\,Y_9$, too. 

Also shown in \reftab{aCP:bench} are the CP asymmetries calculated using a
phenomenological ansatz with $c\bar{c}$-resonances \cite{Kruger:1996cv,Nakamura:2010zzi}.
They are in the general ballpark of the OPE ones,  between the LO and NLO findings, and somewhat smaller than the LO results.

%
\subsection{Untagged CP asymmetries with meson mixing}
\label{sec:untagged-cp-asym}

We consider the decays $B_s, \bar B_s \to \phi (\to K^+ K^-) l^+ l^-$ which
especially for muons are of great importance for hadron collider experiments.
We follow closely \cite{Bobeth:2008ij} to which we refer for details on the full
angular distribution \cite{Kruger:1999xa}.

To account for neutral meson mixing, time-dependent transversity amplitudes need
to be introduced:
\begin{align}
  A^{L/R}_a(t) & \equiv A^{L/R}( \bar B_s(t) \to \phi\,(\to K^+ K^-)_a l^+ l^-)\,, 
  \nonumber
\\
  \bar A^{L/R}_a(t) & \equiv A^{L/R}( B_s(t) \to \phi\,(\to K^+ K^-)_a l^+ l^-)\,,
\end{align}
where $A^{L/R}_a(t), ( \bar A^{L/R}_a(t))$ denotes the amplitude for a meson
born at time $t=0$ as a $\bar B_s$, ($B_s$) decaying through the transversity
amplitude $a=\perp, \parallel, 0$ at later times $t$.

For the time evolution the following parameters which involve the un-mixed
amplitudes at $t = 0$ play an important role
\begin{align} 
  \label{eq:xia}
  \xi^{L/R}_a & = e^{-i \Phi_M} 
  \frac{A^{L/R}_a(0)}{A^{L/R}_a(0) [\delta_W \to - \delta_W]}\,,
\end{align}
where $[\delta_W \to - \delta_W]$ implies the conjugation of all weak phases in
the denominator. Here $\Phi_M$ denotes the phase of the $B_s - \bar B_s$ mixing
amplitude which is very small in the SM, $\Phi_M^{\rm SM}= 2 \arg(V_{ts}^*
V_{tb})$.

The untagged rates $d \Gamma + d \bar \Gamma$ can then be written as
\cite{Fleischer:1996aj}
\begin{align}
  \label{eq:untagged}
  \bar A^{}_a(t) \bar A_{b}^{*}(t) & +  A^{}_a(t) A_b^{*}(t) =
  \frac{1}{2} \bar{A}^{}_a(0)\, \bar{A}^{*}_b(0)
\\
  & \times\left[\left(1 + \eta_a \eta_b\, \xi_a \xi_b^* \right)
  \left(e^{-\Gamma_L t} + e^{-\Gamma_H t}\right) + 
  \left(\eta_a \xi_a + \eta_b \xi_b^* \right) 
  \left(e^{-\Gamma_L t}-e^{-\Gamma_H t}\right)\right] \, ,
  \nonumber
\end{align}
where the chirality indices $L, R$ are suppressed for brevity. Here,
$\eta_{0,\parallel} =+1$ and $\eta_\perp =-1$ are the CP eigenvalues of the
final state and $\Gamma_{L (H)}$ denotes the width of the lighter (heavier) mass
eigenstate of the $B_s$ system.  We also neglect CP violation in mixing, which
is bounded by the semileptonic asymmetry for $B_s$ mesons $|A_{\rm SL}^s|
\lesssim {\cal{O}}(10^{-2})$ \cite{Barberio:2008fa}.

After time-integration follows from \refeq{untagged}
\begin{align}
  \int _0^\infty dt \[\bar A_a(t) \bar A_{b}^*(t) + A_a(t) A_b^*(t)\] & =
  \frac{ \bar{A}_a(0)\, \bar{A}^{*}_b(0)}{\Gamma (1-y^2)} 
  \left[1 + \eta_a \eta_b\, \xi_a \xi_b^* 
  -y \left(\eta_a \xi_a + \eta_b \xi_b^* \right) \right], 
  \label{eq:untagged-tint}
\end{align}
where $\Gamma =(\Gamma_L +\Gamma_H)/2$ and $\Delta \Gamma =\Gamma_L -\Gamma_H$
denote the average width and the width difference, respectively, and $y =
\Delta \Gamma/(2\, \Gamma)$.

Due to the simple transversity structure at low recoil \cite{Bobeth:2010wg}, see
\refeq{Aloreco}, there are only two different time evolution parameters, $\xi_L$
and $\xi_R$, universal for all the $\perp,\parallel,0$ amplitudes. We obtain
\begin{align} 
  \label{eq:xialoreco}
  \xi_{L /R} = e^{-i \Phi_M} 
  \frac{\wilson{9} \mp \wilson{10} + \kappa \frac{2 \hat m_b}{\hat s}\, \wilson{7}
          + Y + \hat\lambda_u Y_9^{(u)} }
  {\wilson[*]{9} \mp \wilson[*]{10} + \kappa \frac{2 \hat m_b}{\hat s}\, \wilson[*]{7}
          + Y + {\hat\lambda_u}^* Y_9^{(u)} }\,.
\end{align}
In the absence of strong phases we find $|\xi_{L /R}|=1$ and in the absence of
CP violation in the rare decays holds $|\xi_{L /R}|=1$  as well.  In the SM, CP
violation in these parameters is very small: $|\xi_{L /R}|-1= {\cal{O}}\left(
  (m_c^2/m_b^2) \, {\rm Im} \, \hat \lambda_u \right)$.

CP-odd observables allow to measure CP violation without tagging the flavour of
the initial $B$ meson (if the asymmetry between $B_s$ and $\bar B_s$ production
is known).  Among the available coefficients in the angular distribution four of
them, $J_{5,6,8,9}$, are CP-odd \cite{Kruger:1999xa, Bobeth:2008ij}. However, in
the low recoil region the coefficients $J_{7,8,9}$ vanish \cite{Bobeth:2010wg},
leaving only $J_{5,6}$ for untagged CP measurements.

For $J_5$ we obtain at low recoil from \refeq{untagged-tint}:
\begin{align}
  \label{eq:untagged-j5}
  \int _0^\infty & dt\,  {\rm Re} 
  \{\bar A^L_{0}(t) \bar A_{\perp}^{L,*}(t) + A^L_{0}(t) A_{\perp}^{L,*}(t) 
        - (L\to R) \}  
  = f_0 f_\perp \times \frac{A_{mix}}{\Gamma (1-y^2)}\,
\end{align}
with
\begin{align}
  A_{mix} & = 2\rho_2 (|\xi_L|^2 +|\xi_R|^2 - 2) + \rho_1 (|\xi_R|^2 -
  |\xi_L|^2) \, .
\end{align}
The formula for $J_6$ is identical after changing the transversity index $0$ to
$\parallel$.

In order to reduce non-perturbative uncertainties, we choose combinations of the
following untagged, time-integrated quantities for normalisation:
\begin{align}
   n_i = \int _0^\infty & dt\, \[|\bar A_{i}^L(t)|^2 + |A_{i}^L(t)|^2 
         + (L\to R) \] 
      = f_i^2 \times \frac{(B_{mix} -  2 \eta_i y\, C_{mix})}{\Gamma (1-y^2)} \, , 
  \label{eq:untagged-norm}
\end{align}
which can be obtained from the angular observables $J_{1,2,3}$. Here,
\begin{align}
  B_{mix} & =
  \rho_1(|\xi_L|^2+|\xi_R|^2+2)+2\rho_2(|\xi_R|^2-|\xi_L|^2) \, , 
\\
  C_{mix} & = \rho_1\Re{\xi_L+\xi_R}+2\rho_2\Re{\xi_R-\xi_L} \, .
\end{align}
Note that $C_{mix}$ only depends on the mixing phase $\Phi_M$.

Normalizing \refeq{untagged-j5} to $\sqrt{n_\perp^{} n_0^{}} $ yields a
mixing-induced analogue of $a_{\rm CP}^{(3)}$:
\begin{align}
  a_{\rm CP}^{mix} & = 
     \frac{A_{mix}}{\sqrt{\(B_{mix}\)^2 - 4y^2 \(C_{mix}\)^2}} \, .
\end{align}
Note that at low recoil $a_{\rm CP}^{mix}$ is insensitive to the sign of $y$.
Simultaneously, the sensitivity to $\Phi_M$ is very low since it enters via
$C_{mix}$ only.  In the limit $y \to 0$  holds
\begin{align} 
  a_{\rm CP}^{mix} & = \frac{A_{mix}}{| B_{mix}|} = -2 \,
  \frac{(\rho_1^2 + 4\, \rho_2^2) \bar{\rho}_2 - 2 \rho_1 \rho_2 \bar{\rho}_1 
       +(\bar{\rho}_1^2 - 4\, \bar{\rho}_2^2) \rho_2 }
       {(\rho_1^2 + 4\, \rho_2^2) \bar{\rho}_1 - 8 \rho_1 \rho_2 \bar{\rho}_2 
       +(\bar{\rho}_1^2 - 4\, \bar{\rho}_2^2) \rho_1} \, .
\end{align}
For $\Delta\rho_i \ll \rho_i, \bar{\rho}_i$ this simplifies further to
\begin{align}
  a_{\rm CP}^{mix} = a_{\rm CP}^{(3)} \, .
\end{align}

The use of the coefficient $J_6$ with normalization to $\sqrt{n_\parallel^{}  n_\perp^{}}$ leads to a second possibility to measure the very same asymmetry  $a_{\rm CP}^{mix} $.

The sensitivity to the $B_s$-mixing parameters $y$ and $\Phi_M$ is low
for realistic values of $y \lesssim {\cal{O}}(0.1)$. We find for  the $q^2$-integrated asymmetries
 that, model-independently,
$| \langle a_{\rm CP}^{mix} \rangle /\langle a_{\rm CP}^{(3)}\rangle -1|$
is below a few percent and hence unlikely  to be measured in the foreseen future.

%
\subsection{CP asymmetries from the angular distribution}
\label{sec:aCP:rel:obs}

Here we summarise the relations between the  low recoil CP
asymmetries $a_{\rm CP}^{(1,2,3)}$ and the CP asymmetries studied previously
in the literature.  We begin by showing various possibilities to extract the
$a_{\rm CP}^{(i)}$ from the angular distribution. We neglect  the small corrections from
finite lepton masses $m_l$ in kinematical factors $\beta_l=\sqrt{1 - 4 m_l^2/q^2}$.

As already mentioned, $ a_{\rm CP}^{(1)} $ equals the total rate asymmetry $A_{\rm CP}$ given as
\begin{align}
  a_{\rm CP}^{(1)} & = 
  A_{\rm CP} = \frac{\Gamma - \bar\Gamma}{\Gamma + \bar\Gamma} \, .
\end{align}
Here,  form factor uncertainties cancel at low recoil.

The asymmetry $a_{\rm CP}^{(2)}$ can be extracted in a multitude of ways from the ratios
\begin{align}
  \label{eq:aCP2:measurement}
  a_{\rm CP}^{(2)} & = 
    \frac{R_{ab} - \bar{R}_{ab}}
         {R_{ab} + \bar{R}_{ab}}, &
       R_{ab}  & = \frac{\sum_a c_a J_a }{\sum_b c_b J_b}, &
  \bar{R}_{ab} & = R_{ab} [J_i \to \pm \bar{J}_i] \, , &  
\end{align}
which provides a cancellation of the form factor uncertainties in the low recoil
region. The ``+'' and ``-'' sign in the third relation apply for CP-even
$i=1,\,2,\,3,\,4$ and CP-odd $i = 5,\,6$  coefficients $J_i$ in the angular decay distribution, respectively. The ratios $R_{ab}$ can have  $a = 5,\,6$ and $b = 1,\, 2,\, 3,\,
4$ and the $c_i$ are auxiliary real numbers.  For example, 
$A_{\rm FB}^{\rm CP}$ can be recovered for
$\sum_a c_a J_a = J_6$ and $\sum_b c_b J_b = J_1 - J_2/3$ (with $J_{1,2}
\equiv 2 J_{1,2}^s + J_{1,2}^c$)
\begin{align}
  a_{\rm CP}^{(2)} & = A_{\rm FB}^{\rm CP} = 
    \frac{A_{\rm FB} + \bar{A}_{\rm FB}}{A_{\rm FB} - \bar{A}_{\rm FB}}\,.
\end{align}

The CP asymmeries $A_i \equiv 2 (J_i - \bar{J}_i)/d(\Gamma + \bar{\Gamma})/dq^2$
and $A_i^D \equiv - 2 (J_i - \bar{J}_i)/d(\Gamma + \bar{\Gamma})/dq^2$ defined
in \cite{Bobeth:2008ij} are related in the low recoil region to the CP
asymmetries $a_{\rm CP}^{(1,3)}$ as 
\begin{align}
  A_3   & = a_{\rm CP}^{(1)} \times \frac{3}{4}\,
    \frac{f_\perp^2 - f_\parallel^2}{f_\perp^2 + f_\parallel^2 + f_0^2} \,, &
  A_4^D & =  -a_{\rm CP}^{(1)} \times \frac{3}{\sqrt{8}}\,
    \frac{f_0 f_\parallel}{f_\perp^2 + f_\parallel^2 + f_0^2} \,, 
\nonumber \\ \label{eq:A3to6}
  A_5^D & = -a_{\rm CP}^{(3)} \times \frac{3}{\sqrt{2}} \, 
    \frac{f_0 f_\perp}{f_\perp^2 + f_\parallel^2 + f_0^2} \,, &
  A_6   & = a_{\rm CP}^{(3)} \times 3
    \frac{f_\perp f_\parallel}{f_\perp^2 + f_\parallel^2 + f_0^2}\ \, ,
\end{align}
whereas $A^{(D)}_{7,8,9}$ vanish. Unlike in $a_{\rm CP}^{(1,2,3)}$ 
in the asymmetries \refeq{A3to6} the form factors do not drop out
and the corresponding uncertainties do not cancel.

The CP asymmetry $a_{\rm CP}^{(3)}$ is identical to the low recoil transversity observables $H_T^{(2)}$ and $H_T^{(3)}$  \cite{Bobeth:2010wg} when
measuring them in an untagged sample containing an equal number of 
$\bar B$ and $B$ mesons
\begin{align}
  a_{\rm CP}^{(3)} & = 
  \begin{cases}
    \displaystyle
    \frac{J_5 - \bar{J}_5}{\sqrt{ -2 (J_2^c + \bar{J}_2^c) 
      \big[2 (J_2^s + \bar{J}_2^s) + (J_3 + \bar{J}_3)\big]}} 
    & \text{for } H_T^{(2)} \\
    \displaystyle
    \frac{J_6 - \bar{J}_6}{2 \sqrt{4 \big(J_2^s + \bar{J}_2^s\big)^2 - 
      \big(J_3 + \bar{J}_3\big)^2}} 
    & \text{for } H_T^{(3)}.
  \end{cases}
\end{align}

The mixing-induced, time-integrated CP asymmetries $A_5^{Dmix}$ and $A_6^{mix}$
defined in \cite{Bobeth:2008ij} are related to $a_{\rm CP}^{mix}$ as follows
\begin{align} \label{eq:A5and6mix}
  A_5^{Dmix} & = -a_{\rm CP}^{mix} \times \frac{3}{\sqrt{2}} \, 
    \frac{n_0 n_\perp}{n_\perp^2 + n_\parallel^2 + n_0^2} \,, &
  A_6^{mix}   & = a_{\rm CP}^{mix} \times 3
    \frac{n_\perp n_\parallel}{n_\perp^2 + n_\parallel^2 + n_0^2}\,,
\end{align}
where the $n_i$ have been introduced in \refeq{untagged-norm}.
As in  \refeq{A3to6}, and unlike in $a_{\rm CP}^{mix} $, the asymmetries given in \refeq{A5and6mix} exhibit a residual form factor dependence.

Ways to extract the angular coefficients $J_i$ from the
full differential decay distribution have been given in 
Ref.~\cite{Bobeth:2008ij}, to which we refer for details and the definitions of the kinematic angles
$\theta_l,\theta_{K^*}$ and $\phi$.
For example,  $J_6$ requires  two bins, $\cos\theta_l \in [-1, 0]$
and $[0, 1]$, while $J_5$ can be extracted from the single-differential
distribution in the angle $\phi$ with additional 
 bins, $\cos\theta_{K^*} \in [-1, 0]$ and $[0, 1]$.

%
%
\section{Model-independent $\Delta B=1$ constraints}
\label{sec:constraints}

\begin{figure}[ht]
\begin{center}
  \subfigure[\label{fig:constraints-c9-c10-loq2+incl}]
    {\includegraphics[width=.45\textwidth]{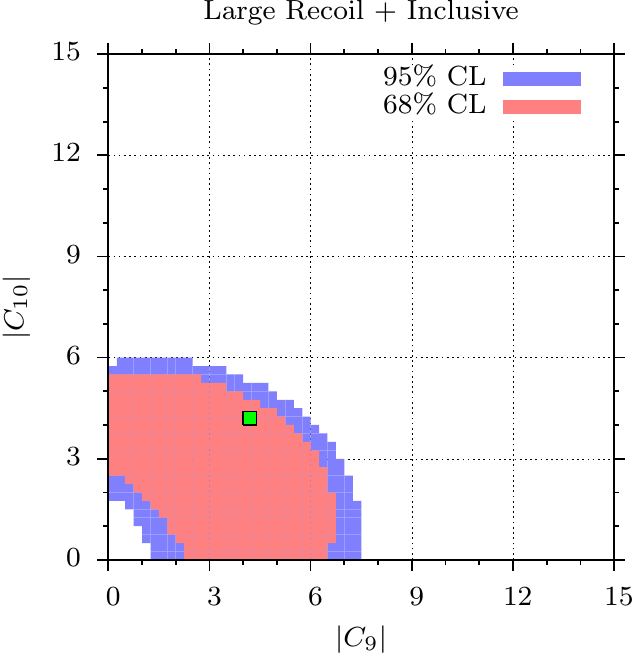}}
  \subfigure[\label{fig:constraints-c9-c10-all}]
    {\includegraphics[width=.45\textwidth]{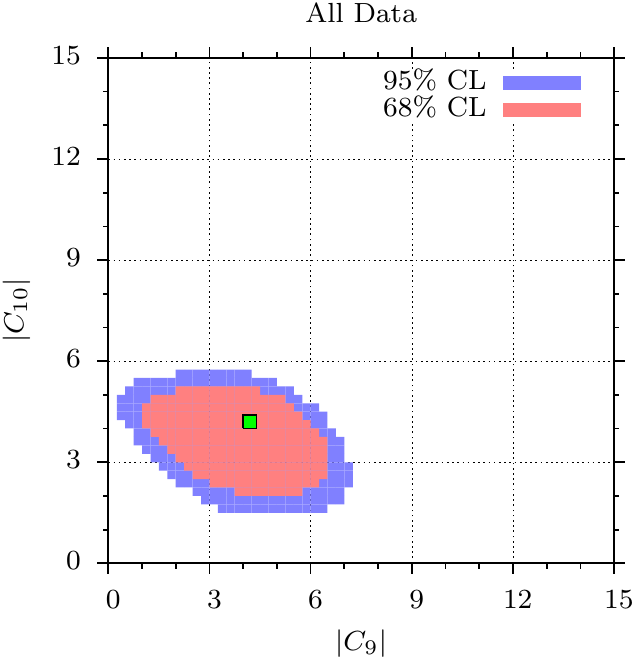}}
\end{center}
\caption{The constraints on $|\wilson{9}|$ and $|\wilson{10}|$ from the experimental
  data as collected in \cite{Bobeth:2010wg}.
  The areas correspond to $68\%$ CL (red) and $95\%$ CL intervals (blue).
  In the left plot  (a), data from the low
  recoil region has been excluded, while the right plot (b) has been obtained using the full
  set of available data. The (light green) square marks the SM.
  \label{fig:constraints-c9-c10}}
\end{figure}

\begin{figure}[ht]
\begin{center}
  \includegraphics[width=.30\textwidth]{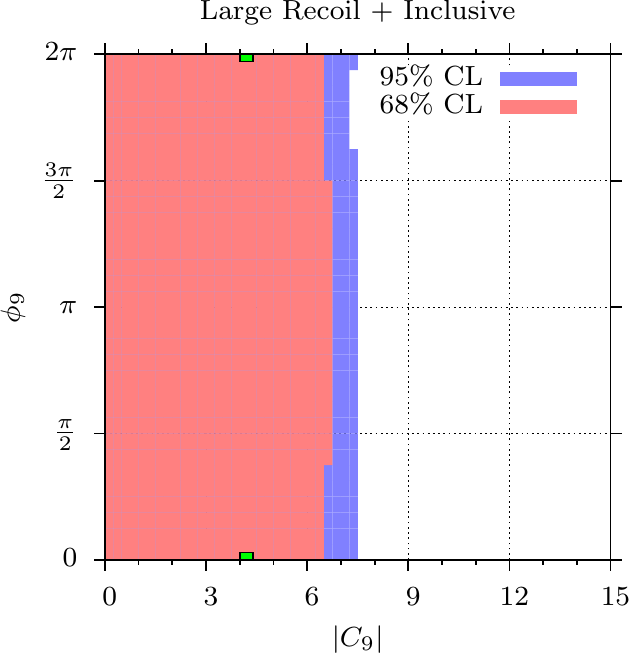}
  \includegraphics[width=.30\textwidth]{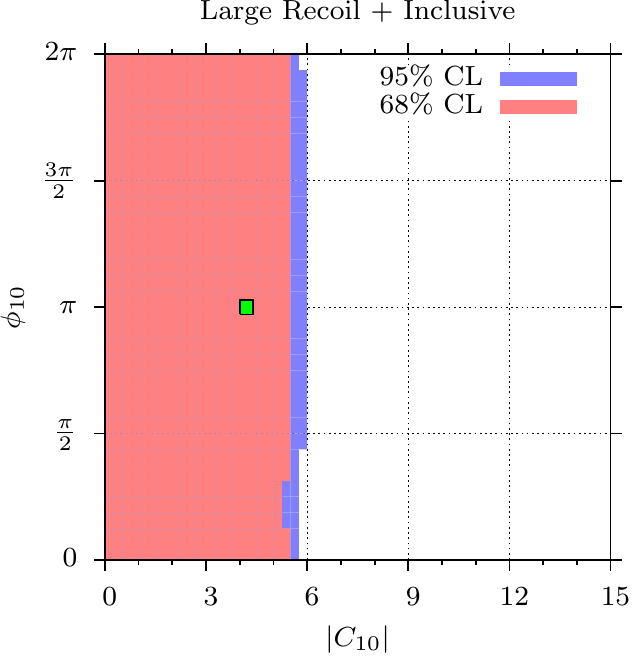}
  \includegraphics[width=.30\textwidth]{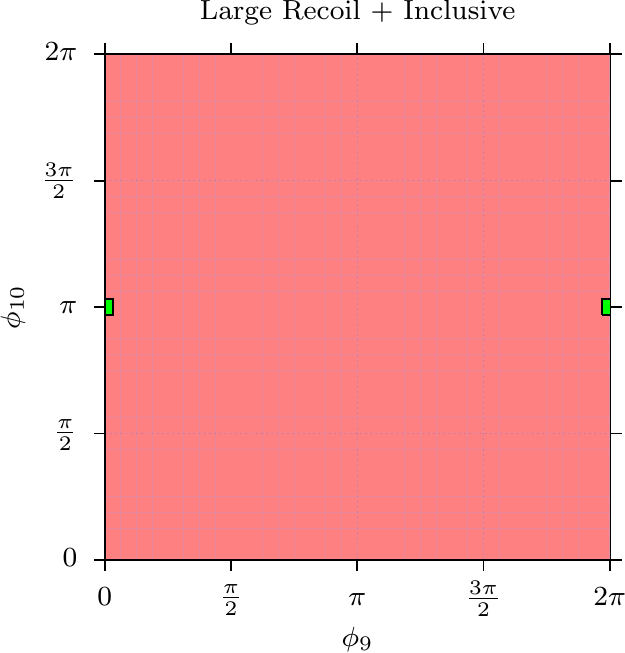} \\
  \includegraphics[width=.30\textwidth]{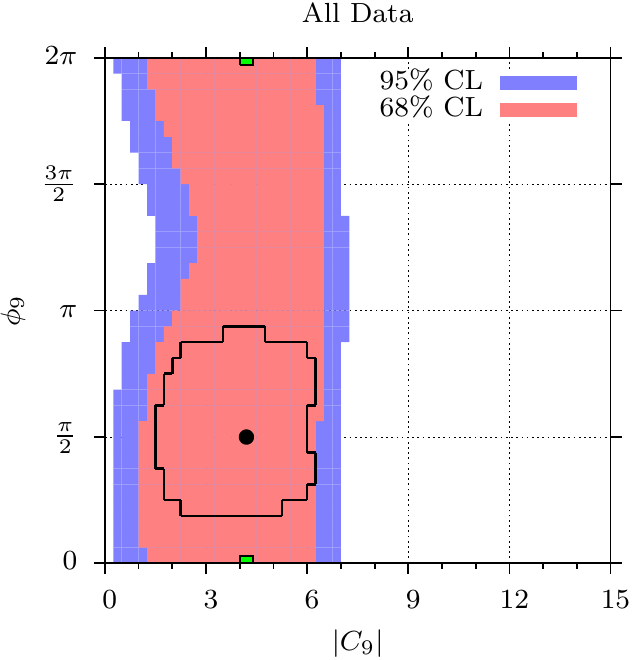}
  \includegraphics[width=.30\textwidth]{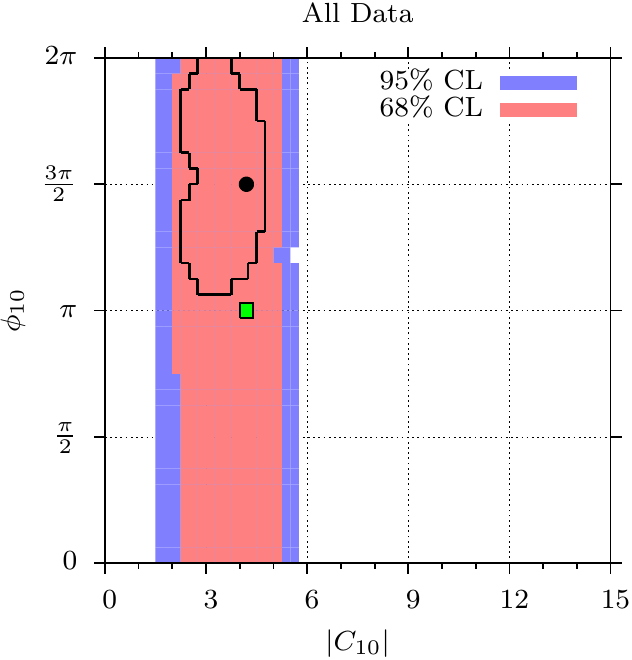}
  \includegraphics[width=.30\textwidth]{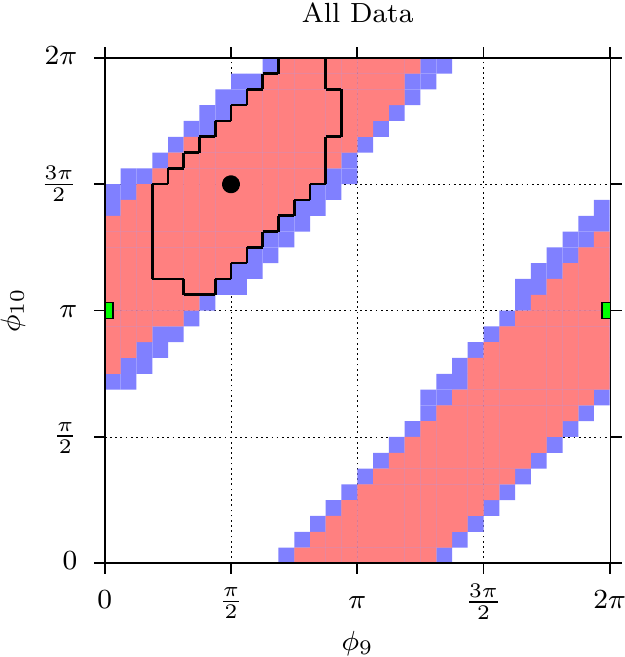}
\end{center}
\caption{The constraints on complex-valued $\wilson{9}$ and $\wilson{10}$ from the
  experimental data as collected in \cite{Bobeth:2010wg}.
  The areas correspond to $68\%$ CL (red) and $95\%$ CL intervals (blue).
  The upper row shows the constraints using the large recoil
  region of the exclusive and inclusive decays only. In the lower row, all data from
  both the low and the large recoil regions are used. 
  The solid (black) lines mark the
  $68\%$ CL regions obtained by employing in addition a hypothetical measurement of
  $\langle a^{(1)}_{\rm CP} \rangle=0.074 \pm 0.01 $ with a central value
  corresponding
  to the one at the BSM benchmark point (\cf~\refeq{benchmark}).
  The (light green) square denotes the SM value of
  $(\wilson{9},\wilson{10})$, while the benchmark point is denoted by the black circle.
  \label{fig:constr-cmplx-c9-c10}}
\end{figure}

We perform a global analysis of the available $b \to s l^+l^-$ decay data in the presence of
BSM CP violation through the Wilson coefficients $\wilson{7,9,10}$, \ie,
allowing them to be complex-valued.
We follow the same approach 
and the same data sources as presented in \cite{Bobeth:2010wg} to perform a
six-dimensional scan of the magnitudes $|\wilson{7,9,10}|$ and phases 
$\phi_{7,9,10} \equiv {\rm  arg}\,\wilson{7,9,10}$. We use the following ranges and binning
\begin{alignat}{4}
    |\wilson{7}|           & \in [0.30, 0.35] & \quad\Delta|\wilson{7}|  & = 0.01\,,
  & \qquad\qquad \phi_7    & \in [0, 2\pi)    & \quad\Delta\phi_7        & = \frac{\pi}{16}\,,\\
    |\wilson{9}|           & \in [0, 15]      & \quad\Delta|\wilson{9}|  & = 0.25\,,
  & \qquad\qquad \phi_9    & \in [0, 2\pi)    & \quad\Delta\phi_9        & = \frac{\pi}{16}\,,\nonumber\\
    |\wilson{10}|          & \in [0, 15]      & \quad\Delta|\wilson{10}| & = 0.25\,,
  & \qquad\qquad \phi_{10} & \in [0, 2\pi)    & \quad\Delta\phi_{10}     & = \frac{\pi}{16}\,.\nonumber
\end{alignat}
The narrow range for $|\wilson{7}| \approx |\wilson[SM]{7}|$ is justified by the
good agreement  of the measured 
$\bar B \to X_s \gamma$ branching ratio with
its SM prediction \cite{Misiak:2006zs,Becher:2006pu}.
For the scan we used and developed further EOS~\cite{eos}, a program for the evaluation of flavour observables.

In order to visualize the constraints, we project the 68\% and 95\% confidence regions of the six-dimensional scan onto the $|\wilson{9}|$--$|\wilson{10}|$ plane,  shown in  \reffig{constraints-c9-c10}.  The projections onto  $|\wilson{9}|$--$\phi_{9}$,
$|\wilson{10}|$-$\phi_{10}$ and $\phi_{9}$--$\phi_{10}$  are obtained likewise,
and are shown in  \reffig{constr-cmplx-c9-c10}.
The data from
$\bar B \to \bar{K}^* l^+ l^-$ decays in the low recoil region provide powerful additional constraints
as can be seen by comparing the results with or without including them.
We find good agreement  between the SM and the data.

The scan procedure returns the allowed ranges, see  \reffig{constraints-c9-c10}, 
\begin{align} \nonumber
  1.0 \leq & |\wilson{9}| \leq 6.5 & (0.3 \leq & |\wilson{9}| \leq 7.3) \, ,\\
  2.0 \leq & |\wilson{10}| \leq 5.3 & (1.5 \leq & |\wilson{10}| \leq 5.8)  \, ,
  \label{eq:c10limits}
  \end{align}
at 68\% CL (95\% CL).  Due to its smallness the above finite lower 95\% CL-bound on $|\wilson{9}|$ is sensitive to the discretisation of the scan, $\Delta |\wilson{9}|=0.25$, and is subject to corresponding uncertainties.

{}From Eqs.~(\ref{eq:SMvalues})  and (\ref{eq:c10limits}) we find  for  the branching ratio  of the decay $\bar{B}_s \to \mu^+\mu^-$ with
respect to its SM value a maximal enhancement by a factor 1.9.  Employing for the decay constant of the $B_s$ meson $f_{Bs}=231(15)(4)$ MeV 
\cite{Gamiz:2009ku}, we obtain the 95\% CL upper limit ${\cal{B}}(\bar{B}_s \to \mu^+\mu^-)< 8 \times 10^{-9}$. The corresponding SM value is given as
 ${\cal{B}}(\bar{B}_s \to \mu^+\mu^-)_{\rm SM} =(3.1 \pm 0.6) \times 10^{-9}$ with the dominant uncertainty stemming  from the decay constant.
   Using  $f_{Bs}=256(6)(6)$ MeV  \cite{Simone:2010zz}, we obtain a slightly larger
upper limit ${\cal{B}}(\bar{B}_s \to \mu^+\mu^-) < 9 \times 10^{-9} $, and 
 ${\cal{B}}(\bar{B}_s \to \mu^+\mu^-)_{\rm SM} =(3.8 \pm 0.4) \times 10^{-9}$.
The SM region of  $\bar{B}_s \to \mu^+\mu^-$ decays will be at least partially accessed by the LHCb experiment with the 2011-2012 LHC run with projected luminosity up to around 2 fb$^{-1}$ \cite{beauty11talk}.

We find the approximate 95\% CL ranges for the phases, see  \reffig{constr-cmplx-c9-c10},
\begin{align}  
  \frac{\pi}{2} \lesssim &\,{\rm arg}\,(\wilson{9}\,\wilson[*]{10}) \lesssim \frac{3\pi}{2} \, ,
\end{align}
with corresponding $2 \pi$-periodic branches. We note that the experimental
information entering our scans stems from CP-conserving data only and  
the constraints on the BSM phases are currently weak. 
While the $B$-factories already measured the rate asymmetry of $\bar{B}\to \bar{K}^*
l^+ l^-$ decays  at the level of ${\cal{O}}(0.1)$ \cite{Barberio:2008fa}, these constraints
are not included in our analysis because they are given for the total
integrated rate only, or are inappropriately binned  such as the 
high $q^2$ rate asymmetry measurement by BaBar, $A_{\rm CP}=0.09 \pm  0.21 \pm 0.02 $, given for $q^2 > 10.24 \GeV^2$ and excluding the $\Psi^\prime$-peak \cite{Aubert:2008ps}.
We checked explicitly that an $\langle a^{(1)}_{\rm CP} \rangle$ measurement at the level of
the latter with theoretical uncertainties taken into account is not significant in the scan.
We find the  values of the CP-observables  $\langle a_{\rm CP}^{(1,3)} \rangle$
to be within -0.2 and +0.2, while $\langle a_{\rm CP}^{(2)} \rangle$ is  unconstrained by  current
data.

To illustrate the impact of a future measurement of the CP asymmetry,
we add hypothetical  data  with a reduced experimental uncertainty  $\langle a^{(1)}_{\rm CP} \rangle=0.074 \pm 0.01 $ to the scan. The central value is inspired by the BSM benchmark,
see \reftab{aCP:bench}.
As can be seen from \reffig{constr-cmplx-c9-c10}, the extended data set
(solid, black lines) adds complementary constraints on the phases, which become
challenging to the SM (green square).

%
\section{Summary}
\label{sec:conclusion}

Building on previous works on CP symmetries  \cite{Bobeth:2010wg}
we identified CP asymmetries $a_{\rm CP}^{(i)}$, $i=1,2,3$ and $a_{\rm CP}^{mix}$ with no leading form factor dependence from the angular
distribution of $\bar{B}\to\bar{K}^* (\to \bar{K}\pi) l^+l^-$ and of $\bar{B}_s,
B_s \to \phi (\to K^+ K^-) l^+l^-$ decays at low hadronic recoil.
The simple amplitude structure following from the heavy quark framework of 
Ref.~\cite{Grinstein:2004vb} in this kinematical region has been crucial in doing so.
We find that the largest uncertainty in these CP asymmetries 
stems from the  renormalisation scale dependence at NLO in $\alpha_s$, which is sizeable, 
followed by subleading $1/m_b$ corrections, see Section  \ref{sec:cp-asym-lorecoil}.

Being strongly parametrically suppressed in the SM,  the CP asymmetries are nulltests of the SM. At the same time the available experimental constraints 
allow for large BSM effects  in the (low recoil $q^2$-integrated)  asymmetries $\langle a_{\rm CP}^{(1,3)} \rangle$ up to  $\sim 0.2$. The asymmetry $\langle a_{\rm CP}^{(2)} \rangle$
is due to its  possibly vanishing normalisation  presently unconstrained.
The mixing asymmetry in $\bar{B}_s,B_s \to \phi (\to K^+ K^-) l^+l^-$, $\langle a_{\rm CP}^{mix} \rangle$  exhibits for realistic values of the 
mixing parameters little sensitivity to the latter and is numerically close to $\langle a_{\rm CP}^{(3)} \rangle$. Note that both $a_{\rm CP}^{mix} $ and $ a_{\rm CP}^{(3)} $ do not require flavour tagging.

By testing the effective theory \refeq{Heff} against the existing data of $b \to s \gamma$ and $b \to s l^+l^-$ decays in the presence of BSM CP violation, we extract the allowed ranges of the Wilson coefficients $\wilson{9,10}$ shown in
 Figs.~\ref{fig:constraints-c9-c10} and \ref{fig:constr-cmplx-c9-c10}.
We find consistency with a related recent analysis
 for real-valued coefficients \cite{Bobeth:2010wg} and with the SM.
 Parameter points with order one deviations from the SM are presently allowed.

To maximize the
exploitation of data we strongly suggest to provide the future  CP symmetries and asymmetries
in $q^2$-bins accessible to systematic theory calculations, such as $1$--$6\GeV^2$ and $\geq 14\GeV^2$, similar to the common binning used in both recent Belle and CDF analyses \cite{:2009zv,Aaltonen:2011cn}.
For completeness we give here  the low recoil
SM predictions for the basic CP-averaged observables, the branching ratio, the forward-backward asymmetry and the fraction of longitudinally polarised $K^*$ mesons, \begin{align}
10^7 \times \int_{14 \GeV^2}^{(m_B-m_{K^*})^2}
  \dd q^2 \frac{\dd \cal {B}_{\rm SM}}{\dd q^2} & =
   2.86 \, {^{+0.87}_{-0.74}}\Big|_{\rm FF}\, {^{+0.11}_{-0.10}}\Big|_{\rm SL}\,
         {^{+0.10}_{-0.19}}\Big|_{\rm CKM} \, {^{+0.08}_{-0.04}}\Big|_{\rm SD} \, ,
       \\[0.3cm]
  \langle A_{\rm FB} \rangle_{\rm SM} & = -0.41\,
                {\pm 0.07}\Big|_{\rm FF}\,
                {\pm 0.007}\Big|_{\rm SL}\,
                {^{+0.002}_{-0.003}}\Big|_{\rm SD} \,,
       \\[0.3cm]
  \langle F_{\rm L} \rangle_{\rm SM} & = 0.35\,
                {^{+0.04}_{-0.05}}\Big|_{\rm FF}\,
                {\pm 0.003}\Big|_{\rm SL} \, ,
\end{align}
respectively. The predictions are based on the improved uncertainty estimate for the subleading power corrections of Appendix \ref{app:sub-lead:crr} and updated CKM input from \cite{Charles:2004jd}, but follow  \cite{Bobeth:2010wg} otherwise.
The largest uncertainty in the above
observables stems from the form factors (FF). Uncertainties smaller than a permille are not given explicitly.

We stress that the transversity observables allow for consistency checks of the theoretical  low recoil framework.  
Since the OPE-breaking corrections generically will spoil the transversity amplitude relation  \refeq{Aloreco} on which
the predictions {\it i)} - {\it iv)} listed in the Introduction  are based on,
the performance of the employed heavy quark framework  can be tested experimentally.

\acknowledgments

We thank Matthew Wingate for useful communication on the lattice results for  $f_{Bs}$
and Frederik Beaujean for advice on  multidimensional analyses.
We are thankful to the technical team of the $\Phi$Do HPC cluster,
without which we could not have performed our scans. D.v.D. is grateful to
Ciaran McCreesh for valuable advice regarding the numerical implementation.
We are grateful to Christian Wacker for uncovering an error in the computation of the confidence regions.
This work is supported in part by the Bundesministerium f\"ur Bildung und
Forschung (BMBF). 

\appendix

\section{Estimate of the subleading power corrections \label{app:sub-lead:crr}}

The subleading power corrections  to 
$\bar{B}\to \bar{K}^* l^+ l^-$ decays at low recoil arise from
the form factor (Isgur-Wise) relations beyond the heavy quark limit  and  the contributions of subleading operators to the OPE \cite{Grinstein:2004vb}.
Both contributions involve the same (three) HQET  form
factors, which are essentially unknown (see \cite{Grinstein:2002cz} for model estimates) but
 in principle are accessible to lattice calculations.

The subleading corrections to the  form factor relations enter the decay amplitudes 
multiplying the small coefficient ${\cal{C}}_7$.
The subleading OPE-corrections  involve
presently unknown Wilson coefficients of order $\alpha_s$. Their knowledge would require a
generalisation of the 2-loop calculation of \cite{Seidel:2004jh} for off-shell
quark states. The latter gives  in general complex-valued results, introducing
new strong phases.
Both corrections can therefore be parameterised as 
\begin{align}
  \label{eq:sub-lead:scal}
 \tilde  r_i & \sim \pm \frac{\Lambda_{\rm QCD}}{m_b} \,
   \left(\wilson{7} + \alpha_s(\mu) e^{i \delta_i}\right) \, , & i=a,b,c \, .
\end{align}
The exact form of the $\tilde r_i$ can be inferred from $\tilde r_i =r_i {\cal{C}}_9^{\rm eff} $, with the $r_i$ given in Ref.~\cite{Grinstein:2004vb}. As  already noted, the
strong phases $\delta_i$ are currently not known.

The transversity amplitudes depend on the $\tilde r_i$ as follows, see  \cite{Bobeth:2010wg} 
for details,
\begin{align}
  A_\perp^{L,R} & = 
  + i \left[(\wilson[eff]{9} \mp \wilson{10}) + 
       \kappa \frac{2 \hat{m}_b}{\hat{s}} \wilson[eff]{7} + \tilde r_a \right] f_\perp \, , 
\\
  A_\parallel^{L,R} & = 
  -i \left[(\wilson[eff]{9} \mp \wilson{10}) + 
       \kappa \frac{2 \hat{m}_b}{\hat{s}} \wilson[eff]{7} + \tilde r_b \right] f_\parallel \, ,
\\
  A_0^{L,R} & = 
  -i \left[(\wilson[eff]{9} \mp \wilson{10}) + 
       \kappa \frac{2 \hat{m}_b}{\hat{s}} \wilson[eff]{7} \right] f_0
  \nonumber
\\
  & -i  N m_B \frac{(1 - \hat{s} - \hat m_{K^*}^2) (1 + \hat m_{K^*})^2 \tilde r_b A_1
           - \hat{\lambda}\, \tilde r_c A_2}
     {2\, \hat m_{K^*} (1 + \hat m_{K^*}) \sqrt{\hat{s}}} \, ,
\end{align}
which mildly breaks the universality of \refeq{Aloreco}. In the numerical
implementation we vary the $\tilde r_i$ and $\delta_i$ for $i=a,b,c$ 
independently within  $ | \tilde r_i |\leq 0.1$ and $\delta_i \in
[-\pi/2, +\pi/2]$ and allow for both signs in \refeq{sub-lead:scal}. 
The  resulting uncertainty  is termed (SL) in this work.
The SM predictions  of some basic CP-conserving observables are given in Section \ref{sec:conclusion}.
Note that this estimation improves on our previous works \cite{Bobeth:2010wg},
where we introduced for each of the transversity amplitudes one
real scaling factor for the corrections to the form factor relations and additionally for each left- and
right-handed amplitude six real scaling factors in order to
estimate the subleading OPE-corrections.

%
%

\end{document}